\documentclass[acmsmall, anonymous=false]{acmart}

\usepackage[utf8]{inputenc}

\setcopyright{rightsretained}
\acmJournal{PACMHCI}
\acmYear{2019} 
\acmVolume{3} 
\acmNumber{CSCW} 
\acmArticle{53} 
\acmMonth{11} 
\acmPrice{}
\acmPrice{}
\acmDOI{10.1145/3359155}

\received{April 2019}
\received[revised]{June 2019}
\received[accepted]{August 2019}

\usepackage[labelfont=bf,textfont=rm]{subcaption}
\usepackage{booktabs}
\usepackage{dcolumn}
\usepackage{natbib}
\def\citepos#1{{\hypersetup{citecolor=black}\citeauthor{#1}}'s \cite{#1}}

\usepackage{color, colortbl}

\usepackage{graphicx}
\usepackage{pdflscape}

\def\BibTeX{{\rm B\kern-.05em{\sc i\kern-.025em b}\kern-.08emT\kern-.1667em\lower.7ex\hbox{E}\kern-.125emX}}

\begin{document}

%
\title[Forensic Qualitative Analysis of Contributions to Wikipedia]{A Forensic Qualitative Analysis of Contributions to Wikipedia from Anonymity Seeking Users}

%

\author{Kaylea Champion}
 \orcid{0000-0001-6196-942X}
 \affiliation{
  \institution{University of Washington}
  \department{Department of Communication}
  \city{Seattle}
  \state{WA}
  \postcode{98195}
  \country{USA}
}
\email{kaylea@uw.edu}

\author{Nora McDonald}
\affiliation{%
  \institution{Drexel University}
  \streetaddress{3141 Chestnut Street}
  \city{Philadelphia}
  \state{PA}
  \postcode{19104}
  \country{USA}
}
\email{nkm39@drexel.edu}

\author{Stephanie Bankes}
\orcid{0000-0002-7947-908X}
\affiliation{%
  \institution{Drexel University}
  \streetaddress{3141 Chestnut Street}
  \city{Philadelphia}
  \state{Pennsylvania}
  \postcode{19104}
  \country{USA}
}
\email{seb397@drexel.edu}

\author{Joseph Zhang}
\orcid{}
\affiliation{
  \institution{Drexel University}
  \streetaddress{3134 Chestnut Street}
  \city{Philadelphia}
  \state{Pennsylvania}
  \postcode{19104}
  \country{USA}
}
\email{jsz37@drexel.edu}

\author{Rachel Greenstadt}
\affiliation{%
  \institution{New York University}
  \streetaddress{6 MetroTech Center}
  \city{Brooklyn}
  \state{New York}
  \postcode{11201}
  \country{USA}
}
\email{greenstadt@nyu.edu} 

\author{Andrea Forte}
\affiliation{%
  \institution{Drexel University}
  \streetaddress{3141 Chestnut Street}
  \city{Philadelphia}
  \state{Pennsylvania}
  \postcode{19104}
  \country{USA}
}
\email{aforte@drexel.edu}

\author{Benjamin Mako Hill}
 \orcid{0000-0001-8588-7429}
 \affiliation{
  \institution{University of Washington}
  \department{Department of Communication}
  \city{Seattle}
  \state{WA}
  \postcode{98195}
  \country{USA}
}
\email{makohill@uw.edu}

%
\renewcommand{\shortauthors}{Kaylea Champion, et al.}

%
\begin{abstract}
By choice or by necessity, some contributors to commons-based peer production sites use privacy-protecting services to remain anonymous. As anonymity seekers, users of the Tor network have been cast both as ill-intentioned vandals and as vulnerable populations concerned with their privacy. In this study, we use a dataset drawn from a corpus of Tor edits to Wikipedia to uncover the character of Tor users' contributions. We build in-depth narrative descriptions of Tor users' actions and conduct a thematic analysis that places their editing activity into seven broad groups. We find that although their use of a privacy-protecting service marks them as unusual within Wikipedia, the character of many Tor users' contributions is in line with the expectations and norms of Wikipedia. However, our themes point to several important places where lack of trust promotes disorder, and to  contributions where risks to contributors, service providers, and communities are unaligned.
\end{abstract}

\ccsdesc[500]{Human-centered computing~Empirical studies in collaborative and social computing}
\ccsdesc[500]{Human-centered computing~Collaborative and social computing}
\ccsdesc[300]{Networks~Network privacy and anonymity}
\ccsdesc[500]{Security and privacy~Pseudonymity, anonymity and untraceability}
\ccsdesc[500]{Human-centered computing~Wikis}
\ccsdesc[300]{Security and privacy~Privacy-preserving protocols}

%
%
%
\keywords{peer production; anonymity; privacy; forensic qualitative analysis; forensic analysis; thematic analysis; Wikipedia; Tor; user-generated content; online communities; threat models}

%

%
\maketitle

\section{Introduction}
In theory, commons-based peer production projects like Wikipedia and GNU/Linux allow for diverse contributions on a global scale. In practice, would-be contributors face widely varying barriers when they seek to participate. Although potential contributors may seek privacy online in order to mitigate perceived threats, such as government oppression or personal harassment \citep{forte_privacy_2017, kang_why_2013}, contributing to peer production projects while maintaining strong anonymity is frequently disallowed \citep{mcdonald_privacy_2019}. 
Are barriers to anonymity seekers' contributions warranted? What kinds of contributions can be expected from anonymity seekers? What happens when anonymous contributors interact with and work beside others in a community where anonymous contributions are often distrusted? 

\begin{figure}[t]
\centering
    \includegraphics[width=0.8\textwidth]{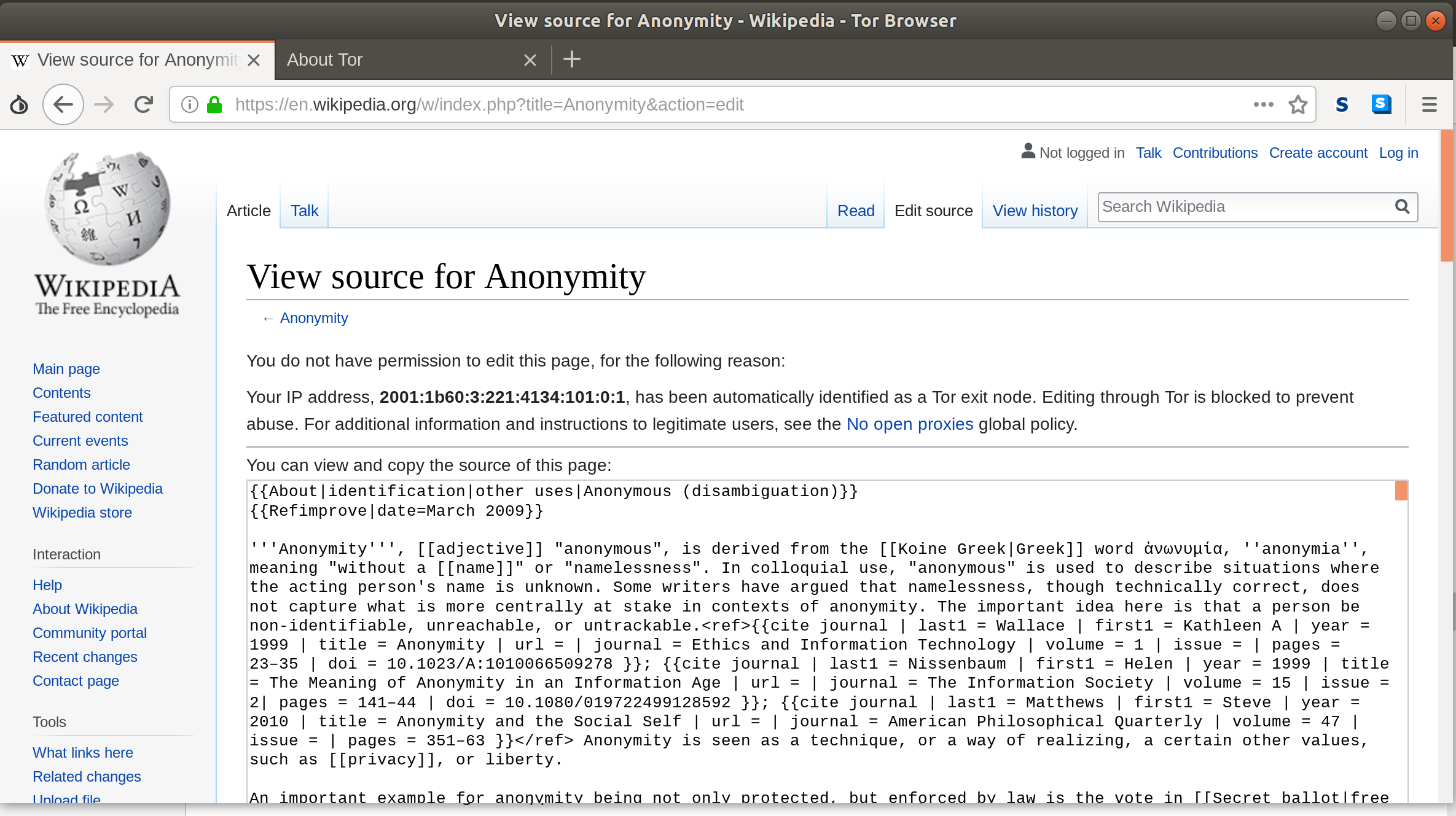}
    
    \caption{When users attempt to edit Wikipedia while using the Tor network, they are presented with a message like the one shown. Users of Tor are told that their IP address ``has been automatically identified as a Tor exit node'' and that ``Editing through Tor is blocked to prevent abuse.'' The concept of exit nodes is described in §\ref{sec:tor}.}
    \label{fig:editBlock}
\end{figure}

To address these questions, we examined contributions to English Wikipedia made through Tor, a secure privacy network that conceals IP addresses and geographic location. Although Wikipedia attempts to block editing through Tor (see Figure \ref{fig:editBlock}), the blocks have sometimes missed Tor addresses or failed to recognize new addresses quickly. As a result, Tor users have managed to edit articles thousands of times \citep{tran_tor_2019}. 
We use these digital trace data as forensic evidence to construct narratives that provide a thick description of contributions to Wikipedia made by Tor users. In turn, we use these narratives as material for a ``contextualist'' thematic analysis \citep{braun_using_2006} that attempts to reflect the limitations of material, question our own assumptions, and maintain awareness of how social context may shape the meaning of what we read and see. 

This paper makes several contributions. First, we describe \textit{forensic qualitative analysis}, an extension of existing qualitative methodologies that we argue can help provide thick descriptions about participants---like anonymity seeking users of Tor---who cannot be interviewed or observed directly but who leave behind rich, if superficially decontextualized digital traces. As our primary contribution, we present the results of a thematic analysis of narratives constructed using this new method on a dataset of contributions made by Tor users to English Wikipedia. We use our knowledge of technology, history, culture, and the Wikipedia community itself to assist us in our interpretation and identify seven themes that suggest editors' intention. Third, we reflect on the challenges of developing trust online, and consider how contribution types carry different risks to contributors and the Wikipedia community. We position these themes with respect to previous work to characterize both motivations to seek anonymity and reasons that may lead service providers and communities to block them.

\section{Background}

In the following sections, we situate our study in the broader literature on why people seek out anonymity online, the means they employ to do so, and the challenges they face from the communities they seek to join. In particular, we consider how anonymity is operationalized and defined in our overlapping empirical settings: the Tor network and Wikipedia.

Our study joins a growing body of work that seeks to understand online anonymity. Although Wikipedia itself characterizes all individuals who contribute without an account as ``anonymous,'' we follow the lead of
previous authors who treat anonymity as a multidimensional concept. For example, \citet{marx_whats_1999} theorizes that anonymity entails obscuring seven types of identifying information:  ``(1) legal name, (2) locatability, (3) pseudonyms that can be linked to legal name and/or locatability...(4) pseudonyms that cannot be linked to other forms of identity knowledge...(5) pattern knowledge, (6) social categorization, and (7) symbols of eligibility/noneligibility'' (p. 100). Online communities often allow people to obscure some types of identity information, while requiring disclosure of others. For example, some communities enforce a ``real name'' policy or require a home address or phone number for verification. Some dimensions of anonymity can be moderated by what someone chooses to share or do within the platform. One can carefully choose a pseudonym to avoid name recognition, decline to disclose personal details, and try to avoid making oneself identifiable through behavior patterns.\footnote{One important caveat here: individuals may be unaware of their `tells'---and given the active work in machine learning and fields such as stylometry, e.g., \citet{brennan_adversarial_2012}, maintaining privacy in this dimension can be extremely difficult.}  Obscuring one's location online typically requires additional effort \citep{forte_privacy_2017}, and some service providers may either deny access or limit the capabilities of individuals who choose not to be locatable \citep{mcdonald_privacy_2019}.

The locatability dimension of privacy reflects an important challenge for Internet users because location information is systematically revealed through IP addresses. An IP address is a unique number used to identify every computer on the Internet. Like a ``to'' and ``from'' address on a piece of mail, IP addresses for both senders and receivers are associated with every piece of traffic sent over the Internet. IP addresses reveal location because they are assigned as part of a larger block to some identifiable and registered unit such as a university, company, or service provider. Using freely available databases, IP addresses can be mapped to approximate address or location by anybody. Network providers can associate individual subscriber homes or even individual computers with the IP address being used, and hence, directly identify the household if not the individual. The EU privacy regulation known as GDPR recognizes the relevance of IP address as a personal identifier 
and requires that it be treated as such \citep{information_commissioners_office_what_2019}. 

Reducing locatability is important to many Internet users because what people post and do online can lead not only to harassment \citep{menking_heart_2015, kang_why_2013}, but also threats to reputation, employment, and harm to self or loved ones \citep{forte_privacy_2017, kang_why_2013}. 
Threats associated with privacy loss may originate from individuals, institutions, or governments, and can have a chilling effect on online expression. Harassed and doxed individuals may be forced off the Internet and into hiding, and journalists and activists may find themselves in jeopardy.
To the extent that anonymous contributors represent minority viewpoints, they may be sources of valuable contributions.

This account of the benefits of privacy online should not be interpreted as suggesting that those who seek privacy online are only those escaping censorship and oppression. Internet users may seek privacy in order to violate laws and norms about free expression or to violate copyright laws \citep{kang_why_2013}. The potentially disinhibiting effects of anonymity and pseudonymity have been linked to increased negative behavior \citep{suler_bad_1998, suler_online_2004, kiesler_social_2009}.
Despite the variety of reasons that Internet users may seek locatability privacy, it is clear that the use of high-quality privacy services is critical to some individuals' ability to participate in public life online and to contribute to peer production projects. We describe one such tool in the next section.

\subsection{IP-Privacy Through Tor}
\label{sec:tor}

Our study concerns users of the Tor network, which protects the privacy of its users' IP addresses. Despite the many reasons that a person might seek anonymity, media accounts of Tor have often described Tor in association with criminal activity \citep{bilton_digital_2013}, or emphasize this type of activity in sensationalist headlines of more nuanced articles \citep{kobie_what_2019, mcgoogan_dark_2016}. Other coverage of Tor reflects a more nuanced view and calls this reputation a matter of ``image'' for a ``useful privacy tool'' \citep{perlroth_tor_2016}. Others describe Tor as an ``internet boogeyman'' that is ``misunderstood'' \citep{menegus_dark_2017}. 
Certainly privacy can also be used to conceal criminal activity \citep{cullum_is_2018}, and multiple studies have sought to measure the extent of illegal material and activity in the Tor network \citep{faizan_exploring_2019,moore_cryptopolitik_2016,owenson_tor_2015}. 

Using Tor represents an explicit decision by a user to employ a privacy tool. 
In conventional network routing, the route---including the IP address reflecting the point of origin---is visible to the recipient of the traffic. Combined with logs from a service provider, access to information about the IP address of the point of origin can allow the unique identification of the location, and often the specific computer, where the traffic originated. By contrast, Tor uses a multi-layered ``onion-routing'' structure that obscures the route to and from, and therefore, the location of the sender. To do this, Tor relies on people worldwide to volunteer machines to act as nodes in the network through which traffic bounces. Each node only knows the step just before and just after it, and no node can see the entire route \citep{huang_onion_2016}.
When someone uses Tor, the places they visit on the Internet can know only the final step in the sequence, known as the Tor exit node. 
Tor dynamically reassigns users to a new exit node IP address as often as every 10 minutes to further obscure the trail back to the user. The list of exit nodes is published and refreshed regularly by Tor.

\subsection{Anonymity and Identifiability in Wikipedia}

Wikipedia allows what is described within the community as ``anonymous'' editing by permitting individuals to contribute without creating an account. These users' contributions are publicly associated with their IP address rather than a username. Although posting publicly as a traceable IP address is not a very effective means of achieving anonymity, this policy lowers barriers to participation for ``newbies'' on Wikipedia \citep{mcdonald_privacy_2019}. Scholars 
have touted the value of such ``anonymous'' contributors,  finding that their work may be more likely to persist \citep{anthony_reputation_2009}. Other work has shown that a significant number of IP-based contributors provide work of high quality \citep{javanmardi_user_2009, anthony_reputation_2009}
and that naive contributions may serve as an indicator of public attention that draws in the efforts of experienced editors \citep{gorbatai_paradox_2014}. 

Anonymity seeking users face a number of challenges on Wikipedia. Contributors without an account are hampered in their ability to accumulate social capital and may be perceived by the community to be less trustworthy.
\citet{oxley_what_2010} observed that contributors to Wikipedia may make six types of authority claims: they may assert their (1) expertise, (2) life experience, or (3) institutional affiliation; (4) use their policy familiarity; (5) cite outside authorities; or (6) leverage social expectations from the community. Of these, we observe that any claim to expertise, life experience, or institutional affiliation would tend to diminish anonymity. 
As a result, they may struggle to successfully assert their position in discussions about what belongs in an article.

\begin{figure}[t]
    \centering
    \includegraphics[width=0.8\textwidth]{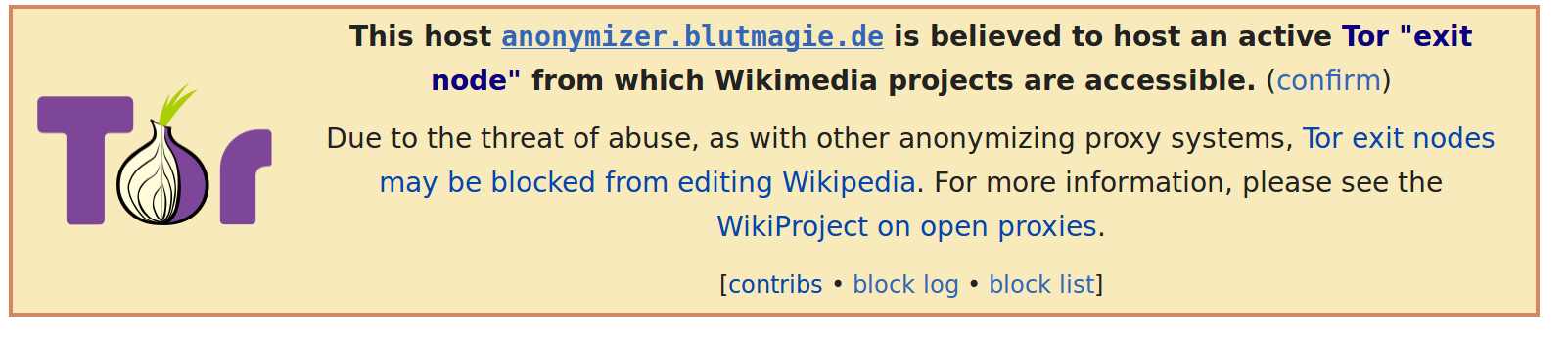}

    \caption{This banner is sometimes placed on user profile pages within Wikipedia to indicate that an IP address has been the source of contributions by Tor users.}
    \label{fig:torBanner}
\end{figure}

Although Wikipedia allows edits from users without accounts, it systematically blocks edits from users of any ``open proxy'' system that seeks to obscure locatability by hiding contributors IP addresses---including Tor. The practical impact of this is that if Wikipedia identifies an address as coming from Tor, which it now does with high speed and accuracy, the individual using that address cannot edit or create an account. Tor users also cannot edit using an existing account unless they first generate a strong positive record of editing without the privacy protection of Tor---a very high bar for anyone to clear if they need privacy protection in order to safely contribute at all \citep{tran_tor_2019}.

Wikipedia blocks systems like Tor because it relies heavily on IP-level blocking to fight spam and vandalism. Because allowing edits from Tor would provide an easy way for users to evade bans, Wikipedia attempts to block all contributions from systems like Tor \citep{tran_tor_2019}.
Once an IP address is identified as part of the Tor network, edits from that address are blocked and profile pages associated with it are marked, as in Figure \ref{fig:torBanner}, so that it is clear that any contributions made from that IP address may have been made by a Tor user.

Given the barriers to Tor editing and the limited anonymity afforded by editing with an IP address, it is difficult to empirically assess what kinds of value anonymity seekers might bring to Wikipedia and similar projects.
We attempt to do so by taking advantage of a unique dataset of edits to Wikipedia made by Tor users that was produced by \citet{tran_tor_2019}.
Tran et al.~take advantage of the fact that although Wikipedia has sought to block Tor since at least 2005, the technology blocking Tor has been imperfect. As a result, over 11,000 edits have been made to Wikipedia using Tor. Tran et al.'s data includes contributions from as far back as 2007 when Tor data became available through 2018. The rate of edits is irregular, and as Tran et al.~speculate, the flaws in Wikipedia's blocking technology may have resulted from multiple factors: delays in the Tor network publishing new nodes, delays in Wikipedia ingesting updated lists of current nodes, irregularities in timing as new Tor nodes joined and left the network, and other timing or stability issues generated by Wikipedia's blocking mechanisms.

In their analysis of Tor users who circumvent the ban, \citet{tran_tor_2019} describe a series of quantitative analyses that suggest that those who contribute to Wikipedia using Tor are similar to other kinds of users, especially those contributing without accounts and new contributors. 
Although a useful first comparison, Tran et al.'s quantitative comparison ignores the content and context of contributions in a way that makes evaluating their value extraordinarily difficult. 
For example, if we examine the details and context of Tor editors' contributions as a Wikipedia community member might, we might recognize ways in which the contributions create unusual risk for, or offering unique benefits to, Wikipedia. 

To help better understand  the potential value and risk associated with anonymity seekers' contributions to social computing systems, we conduct a thematic analysis of Tor edits to English Wikipedia using a novel qualitative analysis approach that we introduce below.

\section{Methodology}

Our methods of reconstructing and interpreting Tor-based Wikipedia contributions comprised several steps. First, we constructed a random sample of 500 edit sessions to English Wikipedia made by users of Tor. We then used this sample to conduct what we call a \textit{forensic qualitative analysis}. This analysis involved examining digital traces within a broader context of other traces as well as making use of our understanding of the community in which they occurred. Next, we conducted thematic coding of the re-contextualized edit sessions by generating, discussing, revising, and applying a set of thematic codes. Finally, we produced a set of detailed narrative memos that described the digital traces, their antecedents, and their effects. This multi-step semantic analysis was conducted iteratively and interactively among a team of four analysts. Although initially session analysis was randomly assigned among the analysts, as narrative threads emerged from our discussion, some sessions were reassigned and an analyst took the lead in exploring the topic in more depth. For example, one edit war spanned multiple pages related to the same conflict about the naming conventions and relative virtues of South Asian schools (see §\ref{sec:thm-editwars}).

\subsection{Sample Construction}
\label{sec:data}

We developed our sample in two stages. First, we grouped edits from the original Tran et al.~dataset of English Wikipedia edits made using Tor into 7,786  edit ``sessions'' \citep{geiger_using_2013}. Second, we drew a random sample of 500 sessions.

This random sampling approach can only yield a sample which is representative of the Tor-based edits that were made despite the block. It is possible that the sample is not representative of the kinds of edits that were attempted or that might have been attempted if Tor were not blocked. For example, the most determined or most savvy Tor-based Wikipedia editors might be overrepresented in the population from which we sampled. Similarly, users who knew that Tor was blocked on Wikipedia might have stopped trying. We discuss this further in §\ref{sec:limitations}.

\citet{geiger_using_2013} define an edit session as a collection of all edits made 
from \textit{the same account} 
to \textit{any page or article} 
as long as there is no more than an hour between edits. We altered this definition 
in two ways and included edits made from \textit{any active Tor node} to the \textit{same page or article} as long as there is no more than an hour between edits. The allowance for multiple Tor IP addresses was made because Tor rotates the exit node in use after 10 minutes.\footnote{\url{https://www.torproject.org/docs/faq.html.en} \textit{archived at} \url{https://perma.cc/75V3-KD4N}} We limit our session to edits made to a single page because considering all edits to all pages would imply that only a single individual using Tor was actively editing on Wikipedia at a given time. 
We chose edit sessions as our sampling unit because, while some contributors to Wikipedia may work for extended periods of time without saving their work, others make multiple subsequent edits, saving repeatedly as they go. The use of sessions, rather than isolated edits, supports the forensic qualitative methodology described in §\ref{sec:forensic}.

Our random sample of 500 edit sessions included 738 individual edits to 438 articles. Some sessions were composed of multiple edits, but most were not. Some articles in the sample were edited in multiple sessions, but most were not. The earliest session was from 2007, and the most recent was from 2017. 

Our ethical commitment to the population of study and the broader community, a challenge noted by \citet{rotman_extreme_2012}, includes the use of only public data that is readily visible to anyone who uses or contributes to Wikipedia. As a result, our work was conducted entirely with logs, article histories, and public IP registrations, and involved no interaction or intervention with research subjects who remained unidentified throughout the process. The research was determined to not be human subjects research by the IRB of the lead author's institution. Despite this, we also pseudonymize names and articles, and in some cases paraphrase quotes, to make reidentification more difficult.

\subsection{Qualitative Forensic Analysis}
\label{sec:forensic}

The distributed and vast nature of online communities can make it difficult to provide the kind of thick descriptions of social phenomena common in traditional ethnographic research \citep{blomberg_reflections_2013, coleman_ethnographic_2010}.

\citet{geiger_work_2010} have advocated combining the analysis of digital traces with participant interviews as part of digital trace ethnography. In empirical contexts like ours, the inability to identify users yields fragmented traces and makes interviewing and the collection of other forms of rich ethnographic data impossible. How can researchers make meaning from traces of people's online activity in settings like ours?
To do so, we conduct what we call \textit{forensic qualitative analysis}. This approach involves attempting to imbue digital trace material with meaning through a detailed study of the trace materials themselves, along with their context and connections with other events and materials. The approach is centered on the experience of the individuals who left the traces and is informed by our own knowledge of the material affordances, behavior patterns, and community values and norms of the empirical setting.

Our approach draws upon innovative methodology used in previous studies in social computing. We are inspired by \citet{twyman_black_2017}, who combined quantitative analysis with careful qualitative reading of discussion pages and historical events to understand interactions of the Wikipedia community with the Black Lives Matter movement. We are also inspired by \citet{nagar_what_2012}, whose careful reading of edit histories was used to observe interactions on Wikipedia as editors participated in collective sensemaking through policy development and interpretation. Finally, we draw from forensic ethnography, which has been used to study corporate crime \citep{van_rooij_toxic_2018}, as well as network forensics, which describes a collection of techniques for reconstructing online events through network traffic analysis \citep{corey_network_2002}. 

\subsubsection{Forensic qualitative analysis in contrast}
Our forensic qualitative analysis methodology builds on others' interpretive and technical strategies but has two unique qualities. First, it directly tackles the problem of the absence of physical participants by inviting the researcher to reconstruct the participant's experience. Second, it is informed by the measures, tools, and strategies employed by community participants who are themselves seeking to interpret the behaviors of others. Wikipedia editors routinely delve into the type of data using techniques similar to our methodology to investigate and interpret the actions of editors when deciding whether to give awards \citep{kriplean_articulations_2008},
when evaluating whether an individual should be made an administrator \citep{burke_taking_2008}, and when investigating complaints, rule violations, or content disputes.\footnote{\url{https://en.wikipedia.org/wiki/Wikipedia:Dispute\_resolution\_requests} \textit{archived at} \url{https://perma.cc/X7V5-UA48}}
Wikipedia is organized to support this type of work. Its public archives include a wealth of data about what changes were made and by whom. User-designed tools for querying this data are shared and hosted on Wikimedia Foundation servers.

Forensic qualitative analysis is particularly useful for studying  anonymous actions.  As \citet{scott_reveal_1998} describes in his theoretical model of anonymous communication: faced with an anonymous author, recipients may seek to reconstruct the identity and intentions of the author based on clues or details contained within the text itself. 

Our method may be useful in other circumstances where research questions concern factors internal to the participant (e.g. motivations and intentions) and where subjects are not identifiable or reachable but where observational digital traces are rich. Examples include discussion boards and chat applications where anonymity is allowed or may even be normative \citep{bernstein_4chan_2011,schoenebeck_secret_2013}. Forensic qualitative analysis may also be useful in contexts where the platform allows participation from individuals both with and without identifiers such as an account.

\subsubsection{Forensic qualitative analysis in Wikipedia}
We began building context for each edit in our sample by looking roughly ten contributions forward and backward within each page's edit history.
For every edit, we attempted to explore Wikipedia's extensive backstage area for additional context. This includes the discussion (i.e., Talk) pages associated with each article. It also included the User and User Talk pages of any editors involved. User pages are often used as personal home pages for Wikipedia contributors and User Talk pages are pages used for interpersonal communication between editors. An individual need not have an account to have a User page or User Talk page.
We went further forward and backward in these edit histories if necessary. 
We examined edits from all editors (registered, IP-based, or Tor-based) in these histories as well as edits made to other articles from the same Tor IP address at about the same time.
We aimed to remain curious for as long as possible, to chase down leads that presented themselves, and to consider other sources within Wikipedia such as block logs and noticeboards. We also used general Internet resources, such as the Internet Archive, the WHOIS IP registration database, and external sites which contributors added links. We sought to reconstruct the timeline and the effects of editing activity.

We tried to gain insight into the state of mind of all the participants. We used the web-based Wikipedia interface that Wikipedia contributors use including the article history page in Wikipedia that allows anyone to navigate all edits to an article in chronological order.
We kept an open mind and considered that vandalism on Wikipedia may reflect mistakes by inexperienced newbies \citep{halfaker_dont_2011, mcdonald_privacy_2019}. To the extent that innocent or unwitting vandalism is part of joining a community of practice, anti-normative behavior may simply be a feature of the learning environment \citep{bryant_becoming_2005, jackson_did_2018}. Other activities that might be labeled as vandalism include the activities from overzealous contributors passionate about a topic.
Our process involved creating detailed notes on each edit session that reflected our attempts to gather context.

Four of the authors conducted an analysis as described above across overlapping sub-samples drawn from our 500 edit sessions. All the authors have experience using Wikipedia. The first author analyzed over 200 sessions, two authors analyzed just under 150 sessions, and a fourth author analyzed 50 sessions. Some sessions were examined by multiple authors as part of our iterative process. In addition to revision IDs and hyperlinks to the archival location of each edited page in each session, we authored metadata including the time and date of edits, indicators of whether edits were reverts and whether the edit was later reverted, and an assessment from the ORES machine learning algorithm that assesses whether an edit might be considered good faith or damaging \citep{halfaker_ores:_2016}. 

Using the sessions themselves and the notes we created while exploring context, we coded our data using an iterative, open thematic coded process as described by \citet{braun_using_2006}.
While coding our data, we met repeatedly to collaboratively generate inductive codes and discuss emerging themes. We eventually arrived at a consensus set of emergent codes and applied these codes to our selection of the data. We discussed definitional issues and questions of interpretation as they arose. Through this iterative process, we assigned one or more of our emergent codes to each edit in our dataset. On some occasions, a single session was composed of different kinds of edits. Once the edits were fully coded, the first and second authors began a process of revisiting our data to write a series of memos that described important themes in our open coding. We returned to our data to create a series of narratives that provided comprehensive descriptions of examples corresponding themes described in our memos. Finally, we grouped these narratives into the seven themes presented below.

\section{Findings}

In this section, we present the most important themes that emerged from our analysis. Each theme reflects a type of contribution made repeatedly by Tor editors to Wikipedia in terms of the intention of the person editing through Tor as well as the reaction of other editors. We observed seven major themes described in sections below: 
quotidian contributions (§\ref{sec:thm-quotidian}), 
bad faith contributions (§\ref{sec:thm-badfaith}), 
activism (§\ref{sec:thm-activism}), 
quality maintenance (§\ref{sec:thm-quality}), 
edit wars (§\ref{sec:thm-editwars}), 
non-article discussion (§\ref{sec:thm-nonarticle}), 
and 
protests against mistrust (§\ref{sec:thm-mistrust}). 
For each theme, we consider potential ways in which the use of a privacy technology may or may not play a role in shaping the different risks faced by communities and individuals.

\subsection{Quotidian Contributions}
\label{sec:thm-quotidian}

The most conspicuous theme that emerged from our analysis reflected the fact that many edits were quotidian in nature. Tor-using contributors frequently engaged in the basic everyday tasks of English Wikipedia and did the same work as other kinds of contributors. These Tor users appear motivated to contribute in the same way many others are: they see a problem they are able to fix or a typo they know how to correct and they hit the edit button in order to do so \citep{bryant_becoming_2005}. Our sense is that these editors may be unaware that their use of Tor to edit Wikipedia is forbidden.

Edit sessions that reflect this theme included Tor-based contributors adding new details to plot summaries of television shows, fixing capitalization errors, and updating the details of bus and train schedules. On the surface, these edits do not appear to be controversial, damaging to the community, or ``high risk'' to an individual.  An example of quotidian contributions that we observed is captured in a one-sentence narrative: 

\begin{quotation}
    On January 31, 2008 at 07:52 UTC, a Tor-based contributor updated the exchange rate for an Asian currency. 
\end{quotation}

\noindent Changing numbers in an article with no explanation and no references might constitute vandalism if the information is incorrect or misleading. A numerical change might go unnoticed, and many articles have only a few citations against which specific facts might be verified. In this case, we conducted a search of historical exchange rates and found records that the exchange rate the Tor editor described was correct to all four decimal places they included. We found no evidence in the contributions made prior to this one that suggested that this edit was anything beyond what it appeared to be. We found no subsequent history to indicate objections or concerns from the community of Wikipedia editors. 

It was easy for us to forget that these contributors chose keep their IP address private by using Tor. As we observed these seemingly mundane activities, we remained aware that we could not be sure if these edits  were only ``quotidian'' to observers with one set of life experiences while controversial to others.  Although we knew that Tor-based contributors were quantitatively similar to other kinds of editors along several dimensions \citep{tran_tor_2019}, we expected that qualitative work would reveal more dramatic differences. Given that people pursue privacy because their perspective places them at risk \citep{forte_privacy_2017}, we imagined that Tor-based contributors would reflect minority viewpoints or making unusual kinds of edits.

Instead, the most important theme to emerge from our analysis is consistent with \citet{anthony_reputation_2009} and \citepos{javanmardi_user_2009} description of persistent, high-quality contributions from people participating without accounts. 
The plurality of the edits we analyzed were quotidian (see §\ref{sec:quant}). In almost all cases, other Wikipedia users gave no indication that they were aware that the Tor edits were made by an anonymity seeking user. In a few cases, an administrator would comment on the fact that the individual was using Tor while removing the edit or banning the user.

We describe the trivial example of the Tor-based edit to a currency exchange rate both to demonstrate the theme and also to offer an example of the level of detail at which our method operates. In an attempt to reconstruct the narrative around an action, we examine the timing, other edits from the same address, other edits before and after, and the content of the edit. We consider our own experience in  Wikipedia to interpret what we observe.

\subsection{Bad Faith Contributions}
\label{sec:thm-badfaith}

Although we found many instances of helpful Tor-based contributors, we also found cases in which Tor users attempted to do harm to the community. Although we identified a range of actions taken in bad faith, the most common forms took the form of vandalism and harassment. As a warning to readers, several of the examples that we describe later in this section include descriptions of violence and self-harm. Examples of harmful contributions from our sample include:

\begin{quotation}
    On March 31, 2011 at 22:02 UTC, a Tor-based contributor replaced the opening sentences of a section of an article about a type of vehicle with the text, ``IM THE BAET DANCER INT UHE WORLD.'' A registered contributor reverted this change 15 minutes later. 
\end{quotation}

\noindent For reasons that should be obvious to anybody that has read an encyclopedia, this text---goofy, all capitals, off-topic, and misspelled---is not helpful or appropriate. In Wikipedia, contributions like this would be called ``vandalism'' and would be quickly and uncontroversially reverted. Other examples of vandalism were those that seemed designed to taunt administrators, set off alarms within the monitoring structures and systems of the Wikipedia community, or to attract attention. For example:

\begin{quotation}
    On November 18, 2011 at 06:00 UTC, a Tor-based contributor updated the User Talk page associated with their Tor IP address seven times over the course of two minutes, with the message ``[name]\footnote{Here, the Tor-based editor typed in the syntax for linking to a user name.} feel like he may commit suicide he needs assistance''. 
    
     The Tor IP address, which had been identified as a Tor node and been blocked on 12 separate occasions since 2006, was blocked again at that point by an administrator who also reverted the Tor-based contributor's edits. 
\end{quotation}

\noindent We cannot judge for certain whether the Tor-based editor was making a sincere cry for help or engaging in vandalism designed merely to provoke others. However, the response of the registered editor suggests that the contribution was understood by other Wikipedians as unproductive. The edit was quickly undone by a registered contributor, who we observe engaged in high-volume vandal-fighting in a process akin to the one described by \citet{geiger_work_2010}. The reverting editor was at the time making many edits per hour and leaving edit summaries that described their work as reverting vandalism. 

We also observed Tor-based contributors engaging in harassment. This sometimes involved adding insults and attacks to other users' User pages. 
While insults posted in articles might constitute vandalism if they do not reference a specific Wikipedia contributor, attacks placed on a User page or the associated User Talk page may be interpreted as an attack on the User page's owner.
Vandalism and harassment in our sample typically resulted in IP addresses being banned from contributing. In many cases, banning corresponded to a discovery by an administrator that the IP address in question was associated with Tor.
We observed examples of harassment as a response to administrators who banned Tor IPs. For example:

\begin{quotation}
    On June 4, 2008 at 02:23 UTC, a Tor-based contributor edited the User Talk page associated with their IP address by inserting an appeal to their having been blocked, with the following text listed in the ``request reason'' portion of the appeal: ``I am [Wikipedia administrator name]. Unblock this IP address, or I will cut off your balls, eat them in front of you so that you an[sic] see it, then chop off your head.'' 
 \end{quotation}  
 
\noindent By placing this threatening comment in an appeal template as a reason for their address to be unblocked, the Tor-based contributor's text was included in a page in Wikipedia dedicated to discussing requests to be unblocked---where it would be read by administrators.

Attacks against hard-working community-selected leaders
suggest a risk posed to communities by contributors using privacy-protecting technology. Although unwelcome, spam, vandalism, and harassment are not unusual in online communities like Wikipedia. Many groups, policies, and technologies exist to counter these types of unwelcome contributions, including artificial intelligence and human moderation of comments and the rapid review of complaints. In Wikipedia, the creation of anti-vandalism automation tools that seek to lessen the burden on vandal fighters is an area of active study and engineering  \citep{asthana_few_2018}. 

\subsection{Activism}
\label{sec:thm-activism}

A third theme we observed reflects community tension of a different kind. Wikipedia's community standards include a requirement that all article text evinces a neutral point of view (NPOV) \citep{reagle_good_2010}.  Establishing neutrality is a difficult and fraught enterprise that was on full display among edits involved in what we call ``activism.''
Maintaining an NPOV involves a constant collective effort and that can draw editors into direct conflict with one another. One person's definition of neutrality may seem a gross distortion of facts to another. Dominant narratives may be widely documented while subaltern points of view struggle to be heard. Some contributors may engage in what is termed ``POV Pushing''---contributions that unfairly or unreasonably bias content in the direction of one's own opinion \citep{reagle_good_2010}.  
A contributor acting as a reviewer of others' content may likewise reflect their own point of view in choosing to support one side or another.
Definitions of NPOV and POV pushing are inherently subjective and a common subject of debate among Wikipedia contributors. For example, people might edit Wikipedia in order to systematically undermine a field of science such as physics or medicine or to change language usage. Other editors may seek to counter these agendas. To a supporter of modern physics or medicine, attacks on those fields may seem like POV pushing, and defending them may be understood as activism. To the skeptic of physics or medicine, casting doubt on these fields likewise may be experienced as activism.

Examples in our sample included a Tor-based contributor who removed the term ``allopathic'' from multiple articles. ``Allopathic'' is described in its Wikipedia article as a pejorative term that supporters of alternative medicine use to describe evidence-based medicine. In this case, the Tor-based editor acted in defense of evidence-based medicine. They were responding to the action of some individual or group which had systematically added the word ``allopathic'' to medicine related articles in order to qualify that the medical knowledge reflected in these articles only reflected one of several legitimate types of medicine.

In another example, a Tor-based contributor updated a politician/lawyer/liberal activist's biographical article to describe him as a ``Democrat politician'' rather than a ``Democratic politician.'' This subtle linguistic shift is part of a larger trend among some supporters of the American Republican Party to describe the Democratic Party as the ``Democrat Party'' in order to distance the party from the adjectival form of the word ``democracy'' \citep{abadi_trump_2017}. 

In another example, we see what could be described as POV pushing morph into activism. Given the tone and timing of the edits, we believe that in this circumstance, a single individual engaged in activism via an IP address then perceived themselves to be at risk and migrated from using an IP address to Tor. Because this is an inductive leap and we may be observing the actions of two individuals, we make note of the evidence in the narrative. 

\begin{quotation}
    On November 14, 2013 at 17:49 UTC, an IP-based contributor edited an article about a mining company. A search of the WHOIS registration of this IP address states that it is registered to a home Internet service provider in a US state. Links in the company article at that time reveal that the company was experiencing financial difficulties and engaged in a conflict with environmentalists who sought to block the company's next project in Europe. The IP-based contributor's addition to the mining company encyclopedia entry accused the company of being complicit with genocide and torture in an Asian country by adding the following text under a new section header with the title ``Support of Genocide and Torture in [Asian Country A]'':
  	
    \begin{quote}
        ``As one of the profitable companies working with [Asian Country B] in the colonization of [Asian Country A], [mining company] is successfully funding and helping to promote genocide in [Asian Country A].  This includes but is not limited to torture (beatings with weapons and steel boots) of [religious figures], rape, imprisonment (with continued torture), electric shock to genitals, relocation and murder.  Psychological warfare also plays an important role as [nationals of Asian Country A] are to be kept far away from mining machinery so profits can continue.''
    \end{quote}
    
    Less than two hours later, someone made an account on Wikipedia with an account name that matched the name of the mining company. This new contributor added updated personnel information and removed both the content added by the activist IP-based contributor and the section of the article describing the company's financial troubles and environmental controversy. The company-named contributor included an edit summary---metadata created by text entered in a box adjacent to the `Publish changes' button labeled ``(Briefly describe your changes).'' Frequently omitted by new users, the use of an edit summary suggests prior experience with Wikipedia. In the edit summary, the company-named contributor claimed that they were ``updating key personnel, subsidiary webpages and deleted section on Genocide in [Asian Country A] which is incorrect information that is not referenced.'' 
    
    Three days later, at 00:33 UTC, 00:35 UTC, and 00:36 UTC, the same activist IP address again updated the mining company page to reference genocide. They inserted text stating that the mining company ``needs love and prayers so that they may overcome this immense greed which does not make them happy.'' They added statements that the company financials included ``funding for an ongoing genocide in [Asian Country A]'' and suggested that environmental approvals were only complete because ``officials and inspectors are paid off.'' 
    
    Then the behavior of the activist IP editor shifts. At 00:45, 00:50, and 00:51 UTC that same day, they remove their own changes. We see edits from a Tor-based editor who adds text back to the same location in the article at 01:02 UTC and removes it again at 01:05 UTC. Although the specific phrasing of the new text is different (``where it funds rape, murder, and torture of innocent [citizens of Asian Country A]'' versus ``[in Asian Country A] helping continue torture, rape and murder''), the substance of the edits is extremely similar. 
    
    Then, between 01:06 and 01:27 UTC, the same Tor-based IP made seven edits continuing the same line of protest. In these later edits, the contributions adopt a more encyclopedic tone. They add encyclopedic statements describing percentage ownership of mines by Western companies and state that the mining company had not responded to a questionnaire from an ethical mining advocacy group. Each of their additions was accompanied by a citation to an external reference. Three days later on November 21, 2013 at 00:23 UTC, the original home IP added text with a reference to an article from Reuters which described the leader of Asian Country B as subject to arrest over genocide allegations if they were to travel abroad.

    A little over two weeks later, on December 10, 2013, at 19:04 UTC, an IP-based contributor deleted the ``Controversies Involving [Asian Country A] Genocide'' section. A search of the public WHOIS registry of this IP address states that these edits came from an IP address registered to the offices of the mining company.
     The company IP-based contributor deleted the ``Controversy'' section and the References section containing evidence for the controversy.
    
    Although no further edits were made from the Tor network at this point, the article received several additional updates. Company-associated IPs and other IPs with no apparent relation to the company removed controversy. Registered contributors and bots applied formatting fixes and added new content that was unrelated to human rights activism. The original home IP used by the activist re-emerged on December 19, reverted changes from the mining company, and manually reinstated some of the same text that the Tor-based editor had added. The activist's changes were again removed by an IP-based contributor, and no further materials about controversies were added to the article over the next year.
   
   This controversy seems to have played out quietly. We were not able to find any arbitration records or Talk page discussion related to the dispute. We do not know why it ended. We do know that by early 2014, Wikipedia's effort to block edits from Tor-based contributors became substantially more successful and the number of edits from Tor-based contributors to the encyclopedia dropped to near zero\citep{tran_tor_2019}.
\end{quotation}

\noindent All sides of this conflict were involved in what Wikipedia might call ``POV pushing'' and ``POV fighting'' and what we call ``activism.'' Both sides violated Wikipedia's rules repeatedly. The initial IP-based contribution that set off the conflict violated Wikipedia norms: its tone was not encyclopedic and it did not include references for its controversial claims.
The edits by people who appear to be associated with the mining company violated many rules as well including Wikipedia's policy on conflict of interest,\footnote{\url{https://en.wikipedia.org/wiki/Wikipedia:Conflict\_of\_interest}  \textit{archived at} \url{https://perma.cc/5J92-NX4U}} Wikipedia's policy against accounts associated with organizations,\footnote{\url{https://en.wikipedia.org/wiki/Wikipedia:Username\_policy\#Shared\_accounts}  \textit{archived at} \url{https://perma.cc/L4BM-FH9Q}} and rules against paid editing of Wikipedia.\footnote{\url{https://en.wikipedia.org/wiki/Wikipedia:Paid-contribution\_disclosure} \textit{archived at} \url{https://perma.cc/Q8LV-SVTM}} 

In this narrative, we infer that the same person was responsible for all the activist-oriented changes critical of the company. It is also possible that two people were working in close collaboration with one another, one using Tor, the other using an IP address. Given textual similarities in their edits and proximities in time, we think the former is more likely. The transition from an IP address to Tor, coupled with removing their own work, is evocative of someone negotiating their approach to a topic, a platform, and their own identifiability. 

This example serves to suggest the kinds of discourse that can be protected or limited by lack of access to privacy enhancing technology. Making accusations and raising awareness of controversy regarding powerful entities like corporations can lead to loss of employment or worse. It may be that the editor who used Tor feared some form of retaliation and was reluctant to continue their activism without protection from a privacy service. 

\subsection{Quality Maintenance}
\label{sec:thm-quality}

Another type of contribution frequently made by Tor users involved the application of English Wikipedia's policies and conventions designed to ensure that articles maintain a standard of quality. This includes both removing materials that violate Wikipedia's policies as well as engaging in collaborative efforts with other contributors.
As Wikipedia has grown, the work of maintaining quality has likewise increased, with a growing proportion of effort going to coordination and upkeep. This a trend was observed as early as 2007 \citep{kittur_he_2007}.

Edits in this theme were typically catalyzed by low quality edits made by others.
When a low quality edit is made to Wikipedia, other contributors can ``revert'' the contribution by undoing it completely, they can try to improve it on their own, and or they can invite collaboration. We found evidence of Tor editors engaging in all three practices. As with quotidian contributions, we saw no evidence that other Wikipedia editors were aware that these contributors were using Tor.

One example of quality maintenance done by Tor-based contributors is the removal of links that violate Wikipedia's external links policy (referred to with the shorthand ``WP:EL''). WP:EL governs what can be placed in the list of links at the end of each article and states that contributors should avoid including links to sites which are, for example, misleading, repetitive, promotional, blogs, social network pages, composed of search results, and, in English Wikipedia, sites which primarily contain non-English content.
We found evidence of Tor-based contributors removing external links that violated this policy that invoked the policy explicitly by including the WP:EL shorthand in their edit summary. 

We found WP:EL invoked by Tor users in edit summaries implicitly and explicitly in a series of edits made to 12 different articles related to religious conspiracy theories and minority religious practices between January 31, 2008 and February 5, 2008. For each article related to the conspiracy theory, at least one of the edits removed a link to the same external website. These edit summaries included:

\begin{quotation}
   On January 31, 2008, at 02:35 UTC, a Tor user edited a biography article about a person allegedly involved in a conspiracy theory with an edit summary that read, ``remove links to self-published personal website; author's qualifications not provided, site may not be reliable''
    
   On January 31, 2008, at 02:38 UTC, a Tor user edited an article referring to a secret cabal with an edit summary reading, ``need a better reference than a spam link to a self-published personal website.'' 
\end{quotation}

\noindent We inspected the website to which links were removed and confirmed that it was a personal page. This is consistent with the statements made by the Tor-based contributor in their edit summaries.  We also found examples of Tor-based edits to the same group of articles adding and fixing reference links and adding the ``\texttt{\{\{fact\}\}}'' tag to an article which causes the phrase ``\texttt{[[Citation Needed]]}'' to appear in a specific place in an article's text. 

We interpret these actions as examples of Tor-based contributors engaged in deliberate efforts to improve the overall sourcing of a group of pages. We observed the use of  authority claims from the ontology suggested in \citet{oxley_what_2010} consistent with maintaining dimensions of anonymity including the use of community social expectations (edit summaries), policies (quoting WP:EL), and external authorities (use of links to sources agreed to be reliable). Subsequently, one of the two Tor IP addresses used by this Tor-based contributor was banned indefinitely (according to Wikipedia policy) by a Wikipedia administrator with the three-word explanation: ``No open proxies.'' 

In other cases, we found Tor-based editors and registered users collaborating to make a stronger contribution than either was able to make alone:
\begin{quotation}
    At 23:13 UTC, on June 20, 2013, a Tor user updated a telecommunications privacy policy article, adding ``However, this has been postponed'' in a subsection discussing the implementation of the policy in a Scandinavian country. The contributor did not include any references or sources for this information. 
    
    Less than an hour later, a registered contributor updated the article to say that the law ``was implemented in 2011...after being postponed'' and added an out-of-date reference. The registered contributor also included an edit summary stating, ``The last information I could find on this is from 2010 (three years ago). The article already cited is from 2011 (two years ago). Please find a better citation if you disagree. A [Scandinavian]-language one is okay.''  
    
    The Tor user made another edit 19 minutes later, stating in the same subsection of the article, ``But this will not be in effect before 1. jan. 2015.'' and adding a more recent, non-English reference. 
    
    23 minutes later, the same registered contributor responded, incorporating what the Tor contributor had added, with an edit summary of: ``Cheers. grammar, date formatting, making sure that language is noted in citation.'' Their edit corrected the grammar of the Tor-based contribution and expanded the metadata for the reference that the Tor-based contributor had provided.
   
\end{quotation}

\noindent In this example, both parties made attempts to signal their collaborative intent. 
The registered contributor used edit summaries to offer explanations for their actions and to give suggestions to their collaborator.
Although the Tor-based contributor did not use edit summaries, the text of their edits revealed that they were responsive to feedback.

Tor-based contributors' familiarity with norms like adding the \texttt{\{\{fact\}\}} tag suggests familiarity with Wikipedia rules and procedures. An experienced editor doing quality maintenance work might choose to protect their privacy for many reasons. For example, they may do so specifically to avoid harassment or stalking. They might also simply be someone who seeks location privacy routinely. Research has documented that those who uphold community policies can face harassment and threats of rape and violence \cite{forte_privacy_2017}.

The editor contributing to a series of articles about a conspiracy theory in our example may have topic-specific reasons to seek to conceal their location and to contribute without an account. In the telecommunications privacy policy article, it might simply reflect the fact that users of privacy technologies like Tor may have expertise in, and a desire to tell others about, privacy-related topics.
These examples suggest that the privacy-seeking community may have value to offer a peer production community by shouldering policy enforcement in circumstances where harassment is a concern, by drawing attention to dubious claims, and through nuanced topical contributions. 

An irony of policy-enforcing contributions from Tor-based editors is the fact that these contributors are themselves policy violators simply by using Tor. According to Wikipedia policy, contributing despite being banned is itself considered grounds for having one's contributions reverted, regardless of the merit or intent of one's contributions.
Although the Tor editors in our sample may not realize this until they find themselves blocked and their contributions removed,  the evidence that at least some of them are familiar with Wikipedia conventions suggests that they may be aware of their contingent access.  

\subsection{Edit Wars}
\label{sec:thm-editwars}

A number of the sessions in our sample that contained multiple edits were examples of what Wikipedians call ``edit wars.'' Edit warring is described by Wikipedia policy as unconstructive back-and-forth editing where two or more editors repeatedly undo each others' contributions. Edit wars are strongly discouraged by Wikipedia policy.\footnote{\url{https://en.wikipedia.org/wiki/Wikipedia:Edit\_warring}  \textit{archived at} \url{https://perma.cc/CW3A-45EG}} 
Edit wars may result in all editing to an article being limited or locked. In some cases, participants in an edit war are enjoined to discuss the dispute first and arrive at some agreement on the article's Talk page before they begin editing again. Some edit wars in which Tor-based contributors participate are resolved in this fashion while others remain contentious. 

We imagine that participants in edit wars would re-tell their participation in a different manner than an outside observer.  Although they might come across as being frustrated, unreasonable, or angry, they may describe themselves as brave, righteous, or trying to uphold fairness. Despite the seeming futility of two people repeatedly undoing each others' work, the circumstance of an edit war are that the first person to step out of the war in essence loses in that the article's text will reflect the update made in the final round. We observed edit wars regarding whether an individual should be listed as an economist, the relative rankings of sports teams in an international contest, the classification of a widely-illegal behavior as a crime, the expansion of an article about a field in physics, and what kind of information should be highlighted in the infobox about characters in a television show. Edit warring often co-occurs with other themes, including activism and harassment, but is characterized by the presence of at least two factions repeatedly undoing each other's work.
 
One example of an edit war in our dataset is a dispute that reflects an offline conflict spilling into Wikipedia. This edit war broke out over a set of articles for several rival South Asian schools. The conflict centered around a registered contributor we will call Cassidy, who engaged in edit warring with one or several people using Tor. 

\begin{quotation}
    We reviewed dozens of edits related to this edit war which first appear in our sample at October 27, 2010 at 02:58 UTC when a Tor user expressed frustration about the contribution patterns of a registered editor. The Tor-based editor placed their complaint on a WikiProject\footnote{Wiki projects are groups of contributors who coordinate their efforts through dedicated group pages.} page corresponding to a South Asian country.
    
    We observed that back-and-forth editing between one or more Tor-based users and Cassidy continued until at least May 5, 2013. The conflict often involved naming of schools, whether the word ``Royal'' should appear in the school name, whether and how to translate school names from a South Asian language to English, and the correct order of words when trying to disambiguate the schools from one another in name. Conflict also arises over which page will host the definition of a sports match between two schools. 
    
    We learn, through Tor-based contributors posting news article links, that this tension is part of some larger conflict that at one point includes violence in the streets between students from two of the schools that are the subject of the edit war.

    Cassidy suffers significant abuse throughout this edit war. On July 3, 2012 at 01:37 UTC,  he is accused of being a ``[unemployed] sick wiki f*ker'' in a comment made by a Tor-based contributor on Cassidy's user page.
    
\end{quotation}

\noindent This multi-article edit war largely comprised a Tor-based editor or editors warring with a single registered editor. The conflict was frequently adjudicated by other registered editors who did not always conclude that Cassidy was acting in good faith or contributing with a neutral point of view. The subject of dispute was difficult for other community members, and for us, to fully comprehend. 

We were able to gather evidence of numerous policy violations on both sides. Although attacks and harassment of Cassidy was clearly an unacceptable norm violation, Cassidy's edits consistently favor a single contested point of view about naming conventions in defiance of guidance from multiple uninvolved contributors.

The fact that Cassidy's attackers had recourse to Tor to conceal their identity may have allowed them to do more damage with their harassment than if they had been identifiable. On the other hand, it may be that the underlying complaint about naming conventions might not have caught the attention of uninvolved Wikipedia contributors without the objections raised by Tor users.
As a registered editor, Cassidy was able to receive direct replies from administrators regarding conflicts and was able to consistently defend his point of view. Tor editors without consistent IP addresses would have likely found it difficult to locate conversations in order to participate in conflict resolution processes.

\subsection{Non-Article Discussion}   
\label{sec:thm-nonarticle}

Another theme described a group of sessions that focused on non-article discussions in Wikipedia. In sessions reflecting this theme, Tor-based contributors participated in activities that range from social chatter, to requests for help with technical problems posed at the Wikipedia Reference Desk, to discussion of policies and misconduct. An example of a narrative that falls into this category is:

\begin{quotation}
  On November 13, 2007 at 07:57 UTC, a Tor user posted an accusation to an administrative board, stating that two Wikipedia administrators were engaged in a form of sockpuppeting called ``strawpuppeting.'' 
  
  The two administrators accused of involvement in strawpuppeting were engaged in a discussion as to whether or not their work to advise marketers on contributing to Wikipedia might represent a conflict of interest.
  
  The Tor-based contributor includes several links to provide circumstantial support for their claim. 
  
\end{quotation}

\noindent The term ``strawpuppet'' is a portmanteau of the ``strawman'' bad faith form of argumentation and the term ``sockpuppeting''. ``Sockpuppeting'' is the practice of creating multiple accounts to be used in bad faith. Sockpuppets might be used to post messages in agreement with oneself or to create the appearance of strong support for a proposal. A sockpuppet might also be used to evade rules that restrict individual contributions, such as the rule that limits the number of times any individual can revert changes to an article in a given day. A ``strawpuppet'' is a sockpuppet which is used not to post support of one's own arguments, but rather to post a strawman attack on an argument which can then be dismantled by the puppeteer. The editor accused of being a strawpuppet was indeed making sweeping claims which disagreed with the two administrators about whether their conduct was appropriate.
While the narrative we developed appears to be a case of a community member calling for accountability, we do not have a view into what follow-up if any occurred. Although this example may be a case of a Tor-based contributor unfairly attacking Wikipedia administrators, legitimate accusations of administrative misconduct may carry risk for identifiable participants. Anonymous individuals may feel more able to criticize community leaders in these ways.

\subsection{Protests Against Mistrust}
\label{sec:thm-mistrust}

One of our examples in §\ref{sec:thm-badfaith} involved a Tor-based contributor responding to being blocked with vandalism and harassment. We also observed instances in which Tor-based contributors protested their treatment by the Wikipedia community while remaining civil. In a series of edit sessions, we saw Tor-based contributors grappling with the mistrust that their contributing without an account generated as well as protesting the policies that prohibit them from contributing. In some cases these protests seem naive and suggest a lack of awareness. In other cases, Tor users appear to be very informed about Wikipedia's policies and processes. 
The following provides an example of a narrative reflecting this theme:

\begin{quotation}
    On June 26, 2011, at 20:28 UTC, a Tor user posted a question in a discussion about whether or not it is forbidden to use Wikipedia's Reference Desk to ask about administrative rulings about individuals. The Tor-based contributor asking this question expanded the inquiry about whether or not this was forbidden in all cases by asking, ``what if someone wanted to ask a question about a notable editor, like for example [name].''[Here, the Tor user invoked the name of a registered editor.]
    
    A registered user responded, ``If you can't ask the question except as a brand-new IP, I suggest the answer is yes.'' [That is, yes, it would be forbidden.]

In the discussion that followed, the Tor-based contributor defended the fact that they asked their question from an IP without an edit history, denied the charge that they were pretending to be a newbie, and described their use of Wikipedia conventions as evidence that they were not pretending to be new.
    
    Other registered editors respond to the Tor-based contributor's question, while yet another asks the Tor-based contributor to give their IP or account name. A registered contributor defends the Tor-based contributor, asking why this information is ``any of [their] business?''
\end{quotation}
    
\noindent This exchange provides insight into the difficulties faced by contributors participating without an account when they interact with others. The absence of a stable identifying account name, the lack of an edit history to which their respondents can refer,  and limited (or non-existent) external cues or relationships to corroborate an initial impression all make productive engagement less likely.

We also observe occasions when Tor-based contributors object to being blocked or banned and try to negotiate access. Two examples include:

\begin{quotation}
    On August 21, 2010 at 23:47 UTC, a Tor user wrote the following using a template that placed their message on a noticeboard for administrators, ``This isn't fair. I am trying to post from the library and I am being blocked.''
    
     On January 28, 2012, at 10:14 UTC, a Tor user posted an unblock request using a template
     with the unblock reason: ``Please unblock this address I would like to create an account.''
\end{quotation}

\noindent Other protests to being banned express confusion:

\begin{quotation}
    On January 20, 2008, at 08:07 UTC, a Tor user writes in their unblock request on their User Talk page, in Chinese, ``What did I do wrong?''[author's translation] The blocked IP in this instance had been blocked since November 12, 2007.
\end{quotation}

\noindent We found no evidence that anybody on English Wikipedia understood the user's request. Other protests make use of Wikipedia-specific understanding: 

\begin{quotation}
    On August 30, 2007 at 01:31 UTC, a Tor user writes in their unblock request on their User Talk page, ``tor should be soft blocked.''\footnote{A ``soft block'' of an IP address in Wikipedia is one in which Wikipedia bans account creation and blocks any edits from a given IP address by individuals who are not logged in, but which does not block editing from that IP for individuals with an account. By contrast, a ``hard block'' does not allow individuals with accounts to contribute from the IP; Tor is hardblocked.}

    15 minutes later, a Wikipedia administrator responds to this request stating, ``There has been no community consensus that the enforcement of \texttt{Wikipedia:No open proxies} is to be overturned yet. While I do sympathize with the situation in mainland China and the Middle East regarding Internet censorship, I have dealt with too many sockpuppets who have abused open proxies in the past and until a technical solution is found, I am not willing to expose this website and its community to such an unacceptable security risk.''
    
    The Tor-based contributor twice posted a response to this denial, including links to comments from a founder of Wikipedia to support their position. These responses were reverted.
\end{quotation}

\noindent In this exchange, we see a lack of trust of the Tor-based editor and a closing-off of further discussion. The administrator's reference to ``sockpuppets'' was a common feature we observed in responses to Tor editors. 

\subsection{Theme Prevalence}
\label{sec:quant}

\begin{table}[t]
  \centering 
  \caption{Prevalence of themes in Tor edit sessions, observed in a sample of 500 sessions. Because multiple codes can be applied to a single observation, percentages do not sum to 100.
} 
  
\begin{tabular}{lcc}
Theme & Number of Sessions & Prevalence\\
\hline
Quotidian & 184 & 37\% \\
Bad faith & 152 & 31\% \\
Activism & 56 & 11\%\\
Quality maintenance & 50 & 10\%\\
Edit wars & 39 & 8\%\\
Non-article discussion & 20 & 4\%\\
Protesting mistrust & 12 & 2\% \\ 
\end{tabular} 
\label{tab:quantThemes}
\end{table} 

The goal of this study is to reconstruct activity through discussion among multiple coders and to interpret the meaning of these activities. It is not to report their frequency with respect to an established standard. As a result, no measure of inter-rater reliability is calculated. We report on the prevalence of themes in Table \ref{tab:quantThemes} to provide requested context for readers but urge readers to interpret these numbers carefully and with some skepticism.  Readers should keep in mind that the ordering of theme prevalence varied somewhat between coders, and session assignment to coders was not fully randomized; although initially assignment was random, when narratives spanned multiple sessions and articles (e.g. the South Asian school edit war), a single coder took the lead in assessing those sessions.

Although each session was coded in terms of at least one of our themes, sometimes multiple codes applied equally well so we applied multiple codes. The application of multiple codes to a single session means the prevalence of all themes do not sum to 100\%. 

Table \ref{tab:quantThemes} shows that the most common theme we observed was quotidian editing which we observed in 184 of the 497 sessions we inspected (37\%). The second most common theme we observed was bad faith, which we observed in 152 sessions (31\%). The third most common theme we observed was activism, in 56 sessions (11\%). 
The remaining themes, in rank order, were: quality maintenance (50 sessions, 10\%), edit wars (39 sessions, 8\%), non-article discussion (20 sessions, 4\%), and protesting mistrust (12 sessions, 2\%). Three sessions in our sample set had been deleted by Wikipedia administrators and removed from the public logs such that they could not be characterized.

\section{Discussion}

The themes that emerged from our forensic qualitative analysis suggest that Tor-based contributors make everyday contributions, they work to uphold quality, they collaborate and argue, they violate norms and misbehave, they grapple with questions of trust and credibility online, and they are drawn into edit wars. In all these ways, Tor editors are like other contributors to Wikipedia. We also found examples where Tor-based contributors introduce specific perspectives that are marginalized or at risk in ways that are consistent with self-protecting anonymity seekers. 
To the extent that policies which block privacy tools discourage contributions from people who might make unique contributions, these policies cause a loss of value to the community.

\subsection{Risks to Anonymity Seekers and the Value to Communities}

\begin{figure}[t]
    \centering
    \includegraphics[width=0.6\textwidth]{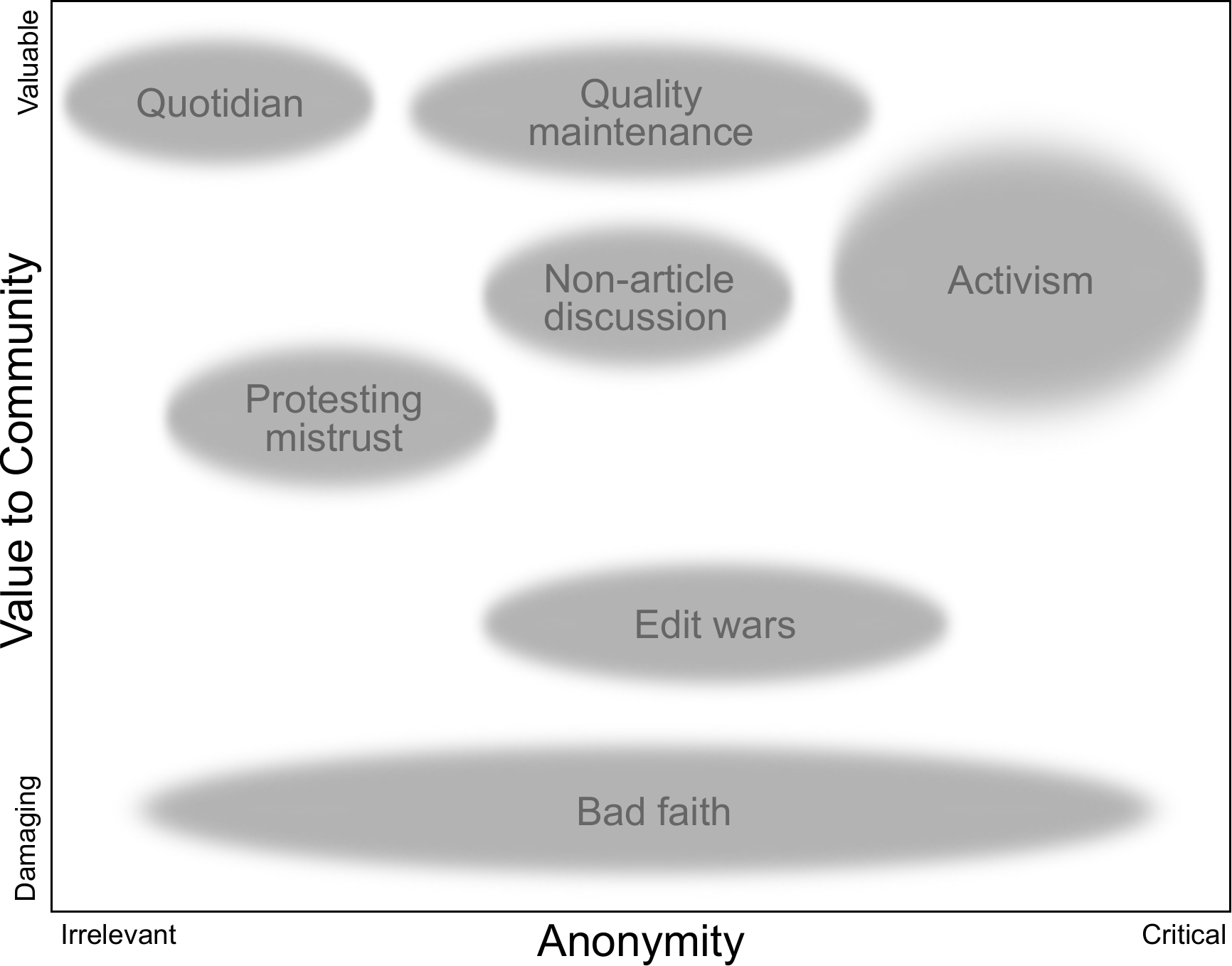}

    \caption{An exploratory mapping of our themes in terms of the value a type of contribution represents to the Wikipedia community and the importance of anonymity in facilitating it. Anonymity protecting tools play a critical role in facilitating contributions on the right side of the figure while edits on the left are more likely to occur even when anonymity is impossible. Contributions toward the top reflect valuable forms of participation in Wikipedia while edits on the bottom reflect damage.
   } 
    \label{fig:bubbleChart}
\end{figure}

We began this paper by discussing the perceived threats that compel would-be contributors to seek anonymity. Throughout our paper, we have shown how Tor users' participation can reflect both value and damage to the Wikipedia community. 
In Figure \ref{fig:bubbleChart}, we explore how the need for anonymity varies in relation to the value brought by anonymity seekers to members in the communities that must decide whether and how to allow anonymous participation. The x-axis in Figure \ref{fig:bubbleChart} attempts to represent each theme in terms of how important a role anonymity plays is facilitating those types of contributions. The y-axis attempts to reflect the value brought to the community if a type of contribution is allowed to occur. The specific locations of the themes are meant to be evocative and we acknowledge that there are valid arguments for many other arrangements.

Although reflecting only an exploratory synthesis, we are confident that edits in our sample occupy all four quadrants. 
Contributions in the top left benefit the community and seem unlikely to require anonymity---though newbies might benefit from such a cloak for trial and error. For example, many of the quotidian edits we observed reflect unambiguous contributions of value to the Wikipedia community and occur frequently among fully identified contributors.
It seems possible that some of these contributions would occur in a context in which Tor was completely blocked. Because they are not placed at risk by these types of contributions, some of the anonymity seeking contributors we observed making mundane edits might have just edited using a pseudonymous account.
Banning contributions from anonymity seekers seems even less problematic for contributions that fall into the bottom right quadrant: they are damaging and seem likely to be at least partially deterred by barriers to anonymous participation. For example, edit wars cause frustration and wasted effort all around. Contributions in this quadrant would likely be welcome casualties of a policy to block anonymous participation. 

Bad faith contributions span the bottom two quadrants. They detract value from the community in that they demand attention and resources to counteract while varying in the risk they pose to an identified bad faith actor. To the extent that using racial slurs or making specific threats online could threaten employment status or result in a police investigation, a strong requirement of identification might reduce some forms of harassment. On the other hand, many forms of bad faith acts, such as the ``BAET DANCER'' instance of goofy vandalism, would pose little risk for the vandal if connected to their identity. Blocking Tor might reduce actions like harassment more effectively than vandalism.
Engaging in non-article discussion is also a complex case. Social discussion and participation in shared governance activities is good for the community in so far as it is pursued in a genuine and transparent manner, but can be dangerous for identifiable individuals. 
However, increased identifiability also places limits on harmful forms of participation, especially manipulations of perceptions of identity such as sockpuppetry and strawpuppetry.

The most important quadrant of the figure is the top right corner which includes contributions that add value to Wikipedia but will not occur if contributors cannot remain anonymous.
For example, many instances of activism we observed fall clearly into this quadrant. Our example of a user documenting the history of notable human rights abuses from a powerful mining company furthers Wikipedia's mission, but places any identified contributor doing so at risk.
Quality maintenance work is likewise supportive of community values and goals but may in some circumstances be triggers for harassment or threats from the individual who authored the removed material.  Previous research has highlighted avoiding harassment as a motivation for using Tor ~\citep{forte_privacy_2017}.
These types of contributions will be less common if barriers to anonymity are kept high. 

Although we offer Figure \ref{fig:bubbleChart} as one conceptual tool to help think through our findings, we put less stock in the specific locations that our themes occupy on the chart. 
When assessing threats and value, we rely on our interpretation, which does not reflect the wide range of risk contexts that may be hard for us to discern from our vantage point as privileged US-based researchers. Some of the world's largest linguistic communities live under various forms of autocracy where edits that seem innocuous or mundane in a democracy with relatively little censorship may present much higher risk. For example, we wondered whether mundane edits we found documenting transportation routes in South Asian countries that are known to censor and surveil the Internet were acts of sedition. We recalled that editing articles about American films or video games might be punishable in oppressive religious regimes. Individuals living in a range of political environments might find themselves under threat from other individuals from whom the law cannot or will not defend them. We examine instead those cases where there is some variation in potential for risk based on the contribution itself.

\subsection{Unique Contributions}

We conducted this analysis with the expectation that we would see some evidence of unique contributions from anonymity seeking contributors, but we did not know what form they might take. We found that even quotidian edits have some potential to be unique to the extent that they represent niche topics. Given that English Wikipedia now contains well over five million pages, there are many specialist topics from which to choose. 
We found unique contributions not only in specialist topics, but also in how privacy seekers may engage with epistemological issues and norm violations. 

We see the strongest potential for unique contributions from anonymity seeking contributors in our activism theme. Activists may place themselves at risk to raise awareness of important issues and may seek to keep their activist work separate from their public persona as a result. Likewise, any community relies on robust debate to maintain accountability and we saw evidence of how anonymity can support productive debate. Journalists cite unnamed sources, newspapers publish anonymous editorials, and ballots are often private. Privacy tools like Tor allow individuals to create privacy for themselves when and where they decide they need and want it. We found that despite rich digital trace data these circumstances are difficult for others---including ourselves---to gauge. 

The decision to block Tor leaves out contributors who require more stringent anonymity including those who contribute activism, quality control, and quotidian edits from locations, or with identities, that put them at elevated risk. We found evidence that these contributions add value that, in some cases, may not be possible to elicit otherwise in that the reflect the perspective of a person at risk. Even edit wars, like those we uncovered in South Asia, may contribute value because they expose an ethnic and (anti-) colonial divide that may not reach western Wikipedia editors otherwise.

\section{Limitations and Future Work}
\label{sec:limitations}

Our work is limited in several important ways.
For example, the way we selected our field site generates certain limitations \citep{blomberg_reflections_2013}. For one, conducting research on individuals who have taken explicit steps to conceal their identity introduces challenges and the potential for error. While we can determine that a given IP address was acting as a Tor node at a given time, we cannot be certain if the edits coming from that node were made using the Tor network or if the Tor node operator was also using the same computer to browse the web directly. We likewise cannot determine definitively whether multiple edits from the same IP are the same person or different Tor users doing similar types of work. We hope that this is mitigated by the fact that we have conducted a careful manual analysis of each edit. We believe that our narrative description would not be substantially altered if a group rather than an individual were responsible for a tightly-spaced series of similar edits.

Our work is also limited in its ability to reflect on what might happen if Wikipedia were to unblock Tor in that the population from which our sample is drawn is unlikely to be representative of people who might want to edit Wikipedia from Tor. For example, regular Tor users who attempted to edit Wikipedia once and received a message explaining that Tor was blocked may never have tried again. In this way, new Tor users might be overrepresented in our sample. In that switching exit nodes repeatedly or requesting specific nodes could make it more likely for Tor users to work around Wikipedia's block, our sample might disproportionately reflect the efforts of savvy Tor users who know how to customize Tor's behavior to take advantage of flaws in Wikipedia's blocking approach. 
If these groups have systematically different interests or goals than the types of people who might edit Wikipedia using Tor in the absence of the block, our analysis might provide limited insight into the question of what might happen in this setting.

Our work is limited in that we have not corroborated our narratives with the community we were observing. We share our interpretation as a laboriously constructed version of what anyone might do when faced with an anonymous message by assembling clues to explain what has occurred. We believe that our narrative-building approach is a practical way to shed light on the behaviors of a hard-to-observe population.
Additionally, it is limited by our decision to sample randomly and then to recontextualize edits as best we could. This means that some elements of long-running narratives, like the South Asian schools edit war, might not be fully captured in our account. As is always the case in qualitative and interpretive research, our account is necessarily partial and incomplete. There are many fascinating narratives that that have unavoidably been omitted.

\section{Conclusion}

Building on a number of existing methodologically approaches, 
we propose a new qualitative methodology---\textit{forensic qualitative analysis}---that extends existing methods to provide us with an understanding of the behavior and intent of people who cannot or will not participate in more direct research techniques.
Using this technique, we contribute to our understanding of a very difficult-to-observe population. We construct our field site around the contributions of individuals protecting their privacy through Tor and the responses of the community to these contributions. Building on prior work that examined the degree to which Tor-based contributors act in good faith and make contributions that are non-damaging, we use our new methodology to establish a series of themes. 

We found that Tor-based contributors make quotidian, good-faith, non-damaging contributions to building the encyclopedia. They use policies to uphold quality and use platform affordances to collaborate with others with varying degrees of success. We also found that they violate policies in ways that damage the quality of the resource: they vandalize, harass, and participate in edit wars. 

Although anonymity seemed incidental or important in many of the types of contributions we examined, it appeared to play a critical in making the contributions possible in others. We saw examples of activists challenging power structures who may have had good reasons for seeking to protect their privacy. Resources which seek to reflect all the world's knowledge may have good reason to embrace those who would champion perspectives outside the status quo that might not be heard in the absence of strong anonymity protection. 

With the deeper look afforded by forensic qualitative analysis, our results suggest that anonymity seekers may add additional value to peer production projects through their work on controversial topics as well as through their ability to challenge prevailing power structures. Our work also suggests that the risks to communities of allowing anonymous contributions may vary enormously. Indeed, different contributors and communities may value particular contributions differently when weighing them against these risks.

We saw many examples of both productive and unproductive engagement through Tor and believe, from the context that were able to gather, that at least some of both types of edits would never had occurred if Tor were blocked completely.
\citet{myagmar_threat_2005} suggest four alternatives for managing risk: accept, transfer, remove, and mitigate. Wikipedia initially accepted participation from Tor before following a strategy of removal with progressively effective techniques. New technology such as automated filtering systems and damage detection capabilities may allow communities to pursue a mitigation strategy in the future. We believe that our findings provide evidence in support of mitigation-based approaches that attempt to maximize value while minimizing damage. We also believe our work can provide some insight into the nuanced ways that value may flow into peer production communities from anonymity seeking users that we hope will inform these approaches in the future.


\begin{acks}
This work was supported by the National Science Foundation (awards CNS-1703736 and CNS-1703049) and included the work of two undergraduates supported through an NSF REU supplement. Feedback and support for this work came from members of the Community Data Science Collective, participants in the Critical and Creative Thinking Studio and in the Science in a Changing World Workshop (both at University of Massachusetts Boston), and from the University of Washington Department of Communication. The manuscript benefited from excellent feedback from several anonymous referees and associate chairs at CSCW.
The project was only possible because Chau Tran generously shared the dataset of Tor edits he had constructed over more than a year of work. The creation of that dataset was facilitated though the use of advanced computational, storage, and networking infrastructure provided by the Hyak supercomputer system at the University of Washington.
\end{acks}

%
\bibliographystyle{ACM-Reference-Format}
\bibliography{bibliography}

%
\appendix

\end{document}